\documentclass[twocolumn]{aastex701}

\usepackage{amsmath}
\usepackage{url}

\usepackage{xcolor}
\definecolor{com_red}{rgb}{.6,.2,0.}
\definecolor{com_turq}{rgb}{.0,.4,0.6}
\definecolor{com_green}{rgb}{.1,.5,.1}
\definecolor{com_gray}{rgb}{.4,.3,.3}

\usepackage{soul}
\setstcolor{com_turq}
\usepackage{comment}

\usepackage{siunitx}

\begin{document}
\title{A test of the \textsc{Dedalus} software for exoplanet atmospheric dynamics}

\author[gname=Rick, sname=Bonhof]{Rick Bonhof}
\affiliation{Kapteyn Institute, University of Groningen, 9747AD Groningen, NL}
\email{q.changeat@rug.nl}

\author[orcid=0000-0001-6516-4493, gname=Quentin, sname=Changeat]{Quentin Changeat} 
\affiliation{Kapteyn Institute, University of Groningen, 9747AD Groningen, NL}
\email[show]{q.changeat@rug.nl}

\author[orcid=0000-0002-4525-5651, gname=James Y-K., sname=Cho]{James Y-K. Cho} 
\affiliation{Martin A. Fisher School of Physics, Brandeis University, Waltham MA 02453, USA}
\email{q.changeat@rug.nl}

\correspondingauthor{Quentin Changeat}



\begin{abstract}
Studying exoplanet flow and variability requires solving atmospheric dynamics equations accurately. 
Here we use the shallow-water equations to evaluate and employ \textsc{Dedalus\,3}, a spectral method-based software package for 
solving differential equations. 
A well-known jet instability test is used for the evaluation; 
then, the package is used to investigate the nonlinear evolution of 
observed, Jupiter's zonal (east--west) jets; 
finally, the package is used to compare hot-Jupiter flows with 
different initial conditions.
Our results indicate that \textsc{Dedalus\,3} can be a useful tool 
for investigating planetary flow dynamics, but careful testing and 
execution are necessary for each problem.
\end{abstract}



\section{Introduction} 

Robust comparisons between simulations and observations are challenging, particularly for exoplanet atmospheres.
They should possess observable variability caused by dynamic large-scale storms and jets, fomented by smaller-scale eddies and waves that lie 
below the resolution of current numerical models and observations.  
Here we utilize the shallow-water equations (SWE) to model the atmosphere
\citep{Pedlosky_1987}:
\begin{eqnarray}
\frac{\partial \mathbf{u}}{\partial t}
 - \nu\nabla^{2}\mathbf{u}
 + g\,\nabla h
 + f\,\mathbf{k}\times \mathbf{u}
 &\ \ =\ \ & -\,\mathbf{u}\cdot\nabla\mathbf{u} \qquad
\\
\frac{\partial h}{\partial t}
 - \nu\nabla^{2} h
 + H\,\nabla\!\cdot\mathbf{u}
 &\ \ =\ \ & -\,\nabla\!\cdot(h\mathbf{u})
\end{eqnarray}
with $\mathbf{u}=\mathbf{u}(\lambda,\phi,t)$ the velocity field; $(\lambda,\phi,t) = $ (longitude, latitude, time); 
$h=h(\lambda,\phi,t)$ the height deviation from the mean layer 
thickness $H$; 
$\nu$ the viscosity coefficient; 
$g$ the gravitational acceleration; 
$f(\phi) \equiv 2\Omega\sin\phi$ the Coriolis parameter; 
$\Omega$ the planetary rotation rate; 
and, $\mathbf{k}$ the local vertical unit vector. 
The SWE is solved using \textsc{Dedalus\,3} \citep{Burns_2020}, a pseudospectral package to: 
{\it i})~validate the package with a well-established 
test \cite[][hereafter G04]{Galewsky_2004}; 
{\it ii})~investigate the nonlinear evolution of observed, Jupiter's 
wind profile; and, 
{\it iii})~study hot-Jupiter flows with different initial states.

\begin{figure*}
    \centering
    \includegraphics[scale=0.57]{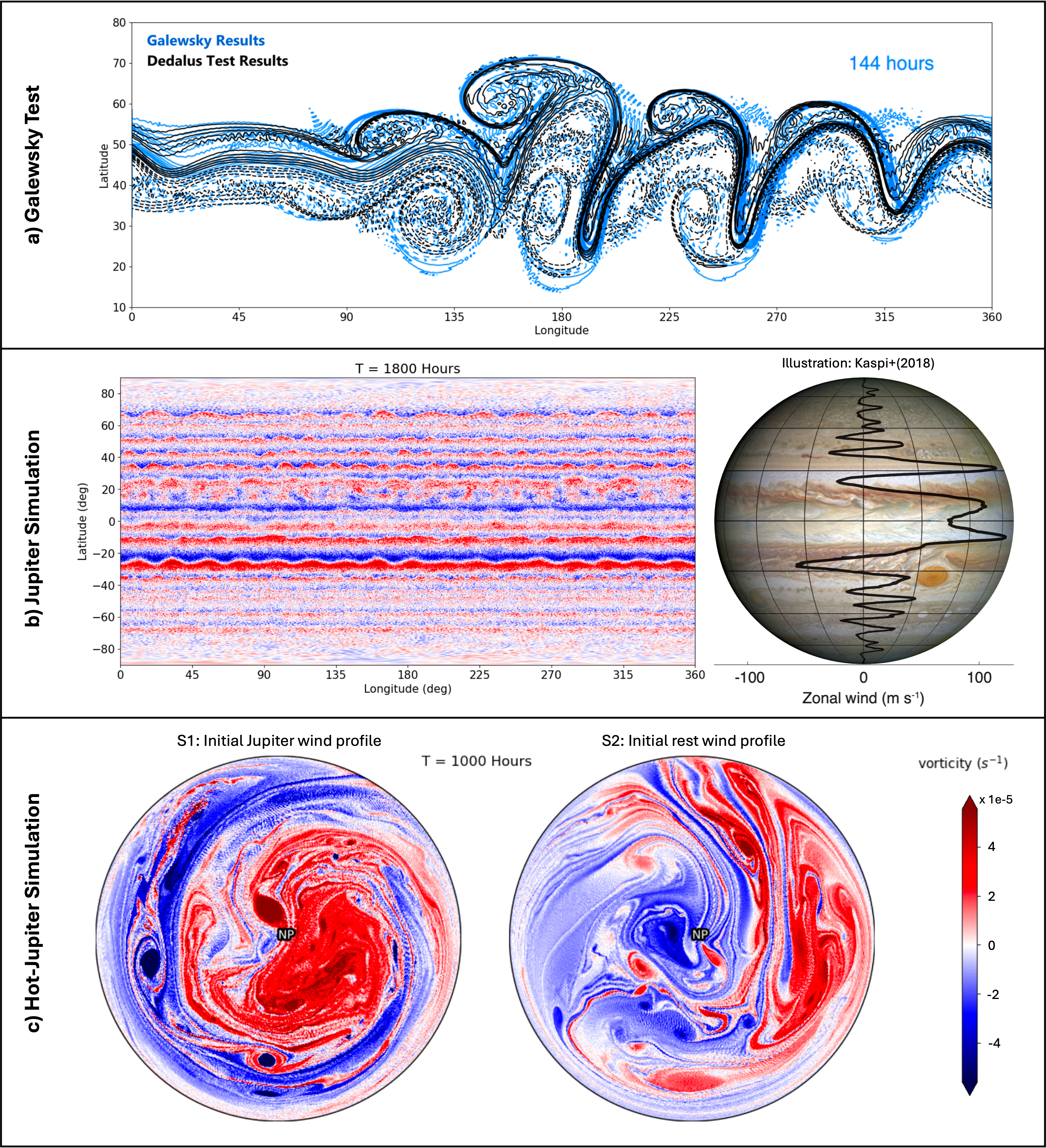}
    \caption{Vorticity fields from T341 simulations. 
    \textbf{a)} \textsc{Dedalus} and G04 with $\nu = 0$ showing 
    2e-5\,s$^{-1}$ contour intervals; 
    calculations match closely but not exactly. 
    \textbf{b)} Jupiter simulation with \textsc{Dedalus} (left) 
    initialized with the JWST zonal wind profile (right); 
    jets and banding are stable on the large-scale.
    \textbf{c)} Polar view of hot-Jupiter simulations with 
    \textsc{Dedalus} (at $\sim$40 ``thermal'' relaxation times),
    initialized with Jupiter's zonal jets (left; S1) and rest 
    (right; S2); the flows showcase the dependence on initial 
    conditions and validate high-resolution simulations.}
    \label{fig:figure}
\end{figure*}

\section{Galewsky et al. Test}

We test \textsc{Dedalus} with the initial-value problem of 
G04---nonlinear evolution of an analytically-specified, 
barotropically-unstable jet at Earth's mid-latitude. 
The jet, in geostrophic equilibrium (horizontal 
pressure-gradient is balanced by Coriolis acceleration), is 
slightly perturbed to initiate dynamical instability.
Following G04, we perform both inviscid ($\nu = 0$) and viscous 
($\nu = 10^5\,\mathrm{m}^2\,\mathrm{s}^{-1}$) simulations at 
\{T42, T85, T170, T341\} resolutions, employing a common 
timestep size ($\Delta t = 30\,$s);
normally, $\Delta t$ is reduced with increasing resolution 
to satisfy the Courant-Friedrichs-Lewy (CFL) 
condition \citep{Strickwerda_2004}.

Our \textsc{Dedalus} results match qualitatively well with 
those of G04 (Figure\,\ref{fig:figure}a); here $\nu = 0$.
For $\nu \ne 0$, the extrema of vorticity 
($\zeta \equiv \mathbf{k}\cdot\nabla\!\times\!\mathbf{u}$) 
at $t = 144\,\mathrm{hrs}$ and height at $t = 4\,\mathrm{hrs}$ 
also match closely with those in G04 (Table~1 therein), differing by
$\delta\zeta_{min} = 5.0 \times 10^{-7}$\,s$^{-1}$,
$\delta\zeta_{max} = 1.3 \times 10^{-6}$\,s$^{-1}$,
$\delta h_{min} = 5.0$\,m, and $\delta h_{max} = 16$\,m
(respectively, 0.68\%, 1.4\%, 0.06\%, and 0.16\% of the G04 
values). 
The smallness of the differences indicates that \textsc{Dedalus} 
is a useful tool for studying planetary flow dynamics with 
the SWE.
However, the differences should not be disregarded---e.g., 
the reduction of fine scale features in Figure\,\ref{fig:figure}a 
suggests unrevealed numerical dissipation or other sources 
(treatment of the nonlinear terms, time-stepping scheme, 
floating-point precision, compiler/library, etc.).
We have explored much of the numerical parameter-space in the 
present study, but further investigation is needed to identify 
the precise origins of the differences and to quantify them.

\section{Jupiter Simulations}

We study the stability of Jupiter's zonal jets  
with \textsc{Dedalus} (Figure\,\ref{fig:figure}b), useful for the 
hot-Jupiter investigation.
We use Jupiter's radius $R_{\scriptscriptstyle J}$, rotation rate, 
and surface gravity, 
as well as $H = 2.7\,\times10^4$\,m (close to the tropopause 
scale-height of Jupiter). 
The simulation is initialized with the JWST zonal jet 
profile \citep{Hueso_2023}, which is again geostrophically balanced 
to obtain an $h$ field consistent with the profile.
A small random stirring of magnitude 100\,m is added to the initial 
$h$ field (only), to abet any instability extant in the jets. 
All the fields are then evolved using equations (1) and (2) 
with $\nu = 0$ and $\Delta t = 30$\,s, 
well below the CFL instability limit.

Figure\,\ref{fig:figure}b shows that the observed jet system is barotropically stable on the large-scale.
The overall banding remains largely unchanged after $\sim$180 
Jupiter days.
This behavior is expected according to linear analysis under the 
rigid-lid (non-divergent) condition:  the necessary condition for instability is that the meridional (north–south) gradient of 
absolute vorticity, $\eta \equiv \zeta + f$, vanishes somewhere in 
the layer \citep{Pedlosky_1987}. 
On Jupiter, the dominant $f$, sharp meridional gradients of 
potential vorticity (closely related to $\eta$), and small 
deformation length help to ``lock'' the zonal jets and bands 
in place.

\section{Hot-Jupiter}

We now present hot-Jupiter simulations using the SWE (Figure\,\ref{fig:figure}c). 
Here the planetary radius is 1.2\,$R_{\scriptscriptstyle J}$, 
$g = 10\,\mathrm{m/s}^2$, and $H = 5\times 10^5\,\mathrm{m}$ 
(for equilibrium temperature of $1500\,\mathrm{K}$).
A forcing term is added to the left-hand side of equation~(2), 
to represent ``thermal forcing'':
\begin{equation}
    (h - h_\mathrm{ref})\,/\,\tau_\mathrm{rad},
\end{equation}
where
\begin{equation}
    h_\mathrm{ref}(\lambda,\phi)\ =\ \Delta h\, 
    \cos\lambda\,\cos\phi
\end{equation}
is the time-independent reference field to which the $h$ field 
is relaxed; $h_\mathrm{ref}$ is the deviation from the mean 
$H_\mathrm{ref}$ (set equal to $H$ without loss of generality), 
and $\Delta h = 0.1 H$; $\nu = 10^5\,\mathrm{m}^2\,\mathrm{s}^{-1}$ and $\Delta t = 30$\,s.

Simulations S1 and S2 (Figure\,\ref{fig:figure}c, left and right respectively) show the flow at $t = 40$~planet days, evolved 
from different initial states: S1 from the Jupiter's zonal wind 
profile and S2  from rest. 
The flow differs markedly.
This is consistent with \cite{Thrastarson_2010}, and expected 
from the difference of initial energy content.
S1 shows a cyclonic ($\zeta > 0$) polar vortex displaced from the 
pole, with many secondary storms; 
most storms are dynamic and long-lived.
The flow is strongly steered by the remnant of the initial jets, 
completely destabilized by the imposed forcing. 
In contrast, S2 shows a smaller, anti-cyclonic  ($\zeta < 0$) 
polar vortex closer to the pole, with more (generally weaker) 
secondary storms. 
Notably, both simulations have polar and equatorial flows similar 
to that of \cite{Cho_2003}---{\it a}zonal and variable.   
In the absence of the initial random stirring (S2), meridional 
symmetry across the equator is almost perfectly preserved.

\section{Conclusions}

Solving the SWE, we have tested \textsc{Dedalus\,3} via a 
well-known test case---with and without explicit 
dissipation. 
The package appears to be a useful tool for investigating 
planetary flow dynamics. 
However, small but noticeable differences are seen, suggesting 
that further investigations are warranted to bolster the 
preliminary conclusions here and highlighting the importance 
of careful validations for exoplanet studies in 
general \citep{Polichtchouk_2014,Cho_2015,Skinner_2021};
quantitative differences can arise even when similar methods 
(here spectral) are employed.
Sample \textsc{Dedalus} simulations of cold- and hot-Jupiter 
flows are also provided, showing the gross stability of Jupiter's 
zonal jets obtained with JWST and providing additional 
confirmation of past hot-Jupiter atmospheric dynamics simulations with high 
resolution \citep{Cho_2003,Thrastarson_2010}. 


\vspace*{.5cm}

{\it Software:} \textsc{Dedalus3}, \url{https://dedalus-project.org/} \\
    
{\it Acknowledgment:} NWO Grants VI.Veni.242.091 and 2024.034.

\bibliography{main}{}
\bibliographystyle{aasjournalv7}



\end{document}